\documentclass{aastex}
\usepackage{spr-astr-addons}
\usepackage{textcomp}
\usepackage{amssymb}
\usepackage{dsfont}
\usepackage{leftidx}

\newcommand{\bo}{\ensuremath{\boldsymbol{B}_0}}
\newcommand{\erf}[2][]{\ensuremath{\text{erf#1}\!\left(#2\right)}}

\newcommand{\kB}{\ensuremath{k\!_B}}

\newcommand{\Rl}{\ensuremath{R_{\mathrm L}}}

\newcommand{\be}{\begin{equation}}
\newcommand{\ee}{\end{equation}}
\newcommand{\bs}{\begin{subequations}}
\newcommand{\es}{\end{subequations}}

\newcommand{\R}{\ensuremath{\mathds{R}}}
\newcommand{\C}{\ensuremath{\mathds{C}}}

\newcommand{\pa}{\ensuremath{_\parallel}}
\newcommand{\se}{\ensuremath{_\perp}}

\newcommand{\Om}{\ensuremath{\varOmega}}

\newcommand{\Ga}{\ensuremath{\varGamma}}

\newcommand{\uint}{\ensuremath{\int_{-\infty}^\infty}}
\newcommand{\Ff}[1]{\ensuremath{\leftidx{_2}{F}{_2}\!\left({\displaystyle #1}\right)}}

\newcommand{\f}[1]{\ensuremath{\boldsymbol{#1}}}

\newcommand{\dd}[2][]{\ensuremath{\frac{\mathrm{d} #1}{\mathrm{d} #2}}}
\newcommand{\df}{\ensuremath{\mathrm{d}}}

\begin{document}
\title{Interstellar Turbulent Magnetic Field Generation by Plasma Instabilities}

\shorttitle{Magnetic Field Generation}
\shortauthors{Tautz \& Triptow}

\author{R.\,C. Tautz} \and \author{J. Triptow}
\affil{Zentrum f\"ur Astronomie und Astrophysik, Technische Universit\"at Berlin, Hardenbergstra\ss{}e 36, D-10623 Berlin, Germany}
\email{rct@gmx.eu}

\begin{abstract}
The maximum magnetic field strength generated by Weibel-type plasma instabilities is estimated for typical conditions in the interstellar medium. The relevant kinetic dispersion relations are evaluated by conducting a parameter study both for Maxwellian and for suprathermal particle distributions showing that micro Gauss magnetic fields can be generated. It is shown that, depending on the streaming velocity and the plasma temperatures, either the longitudinal or a transverse instability will be dominant. In the presence of an ambient magnetic field, the filamentation instability is typically suppressed while the two-stream and the classic Weibel instability are retained.
\end{abstract}


\section{Introduction}

Galactic magnetic fields are ubiquitous \citep[see][for an overview]{bec96:mag}. Even in galaxies at high redshifts, magnetic fields have been found \citep{ber08:gal}. The correlation between far-infrared radiation of massive stars and radio emission produced by synchrotron radiation of energetic particles in the surrounding magnetic fields \citep{mur09:ska}. Such implies a connection between the formation of massive stars and galactic magnetic fields.

The generally accepted model for the generation of galactic magnetic fields is found in the dynamo process \citep{bec96:mag,bra05:dyn,kul10:gal}, which, however, requires a seed magnetic field \citep{sch05:ori,sch12:amp}. Among other processes \citep[e.\,g.,][]{ryu12:lss,dur13:cos}, seed fields can be generated by plasma instabilities for example in the neighborhood of massive stars that ionize the surrounding interstellar medium \citep{sch12:equ}. A special class of such instabilities generates modes that purely grow in time and do not propagate---the so-called ``aperiodic'' modes \citep{wei59:wei,tau12:rad}. The fact that such modes can be emitted spontaneously even in unmagnetized plasmas \citep{yoo07:spo,tau07:spo,yoo12:sp1,laz12:sp3} again underscores the validity of the process. On smaller scales, magnetic fields play an important r\^ole in the formation of molecular clouds \citep{ino12:mol}, star formation, and thermally unstable interstellar flows \citep{man12:dyn}. Furthermore, aperiodic modes are essential for particle acceleration at cosmic shocks \citep[e.\,g.,][]{rev08:tra,nie10:bea}. In general, the coupling of matter and magnetic fields is confirmed by the typical scaling $B\propto\sqrt n$ for relatively high particle densities \citep{hei05:mol}.

Because of the typically low plasma densities, the relevant processes have to be described using kinetic plasma theory \citep[see, e.\,g.,][for an introduction]{dav83:kin,rs:rays,tau12:nov}, which has a long tradition. Much of the progress in cataloging waves in plasmas, both non-relativistically as well as relativistically, has ably been summarized by \citet{cle69:ele} and, with astrophysical applications much to the fore, by \citet{rs:rays}, where copious references to the many advances in understanding such waves are to be found. Typically, concentration is focused on simplified geometries, for example modes propagating parallel or perpendicular with respect to a given symmetry axis such as a streaming direction or an ambient magnetic field. A considerable amount of work has been done on oblique propagating wave modes \citep[e.\,g.,][]{bre06:obl,gre07:bpi} and general coupling effects between the various modes \citep[e.\,g.,][]{tau06:is1,tau07:har,tau12:3x3}.

Here, the maximum magnetic field strength generated by a special class of plasma instabilities---so-called ``Weibel-type'' instabilities \citep{wei59:wei,fri59:wei,ach07:we1,tau12:rad}---will be investigated by means of a parameter study. Such instabilities have been the focus of intense research for some time regarding both their linear and non-linear stages. Based on a recently developed method \citep{tau11:max}, the maximum growth rate and the associated wavenumber can be efficiently determined. In contrast to the work mentioned above, which explained the generation of seed magnetic fields in the early universe, we aim at small-scale magnetic field generation using present-day conditions. Instead of a fully non-linear calculation \citep[see, e.\,g.,][]{ach07:we2}, a simple estimation will be used that has been confirmed by numerical particle-in-cell simulations. Throughout, Gaussian cgs units will be used.

\begin{figure}[tb]
\centering
\includegraphics[bb=97 282 487 559,clip,width=\linewidth]{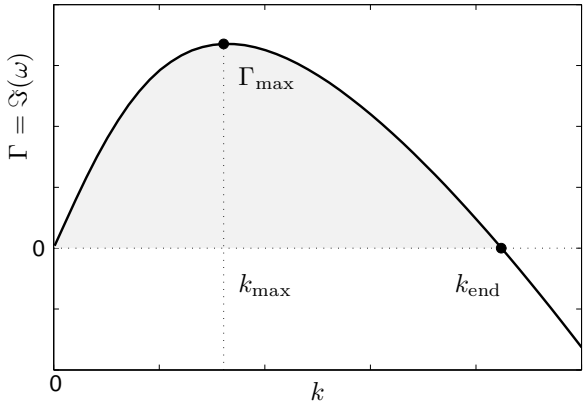}
\caption{Schematic plot of the imaginary frequency part---the growth rate, \Ga---as a function of the wavenumber in arbitrary units. The maximum growth rate, $\Ga_{\text{max}}$, lends its name to the associated wavenumber, $k_{\text{max}}$. The wavenumber at which the growth rate vanishes, in contrast, is labeled $k_{\text{end}}$. The shaded area denotes the region of instability.}
\label{ab:growthrate}
\end{figure}

This article is organized as follows: In Sec.~\ref{tech}, the dispersion relation are introduced together with the initial distribution functions that describe the streaming and temperature anisotropies. The relations used to estimate the maximum magnetic field strength and spatial scales on which it varies are introduced in Sec.~\ref{estim}. In Sec.~\ref{param}, a parameter study is conducted to illustrate the magnetic field growth for various combinations of the parameters for particle density, electron temperature, and streaming velocities, which all form the anisotropy pattern responsible for the instability behavior. Additionally, the effect due to a suprathermal particle population will be demonstrated. Sec.~\ref{summ} provides a short summary and a discussion of the results.

\section{Technical Development}\label{tech}

\begin{figure}[tb]
\centering
\includegraphics[bb=97 282 487 559,clip,width=\linewidth]{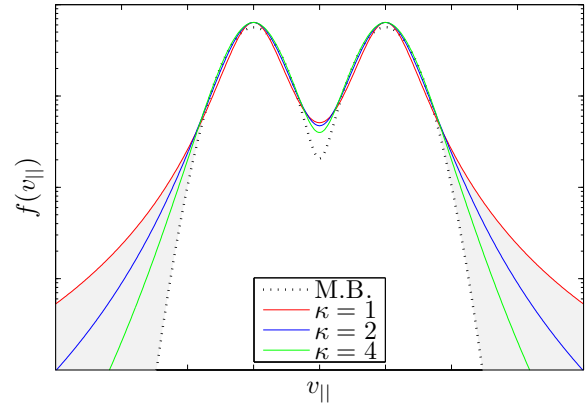}
\caption{(Color online) The Maxwell-Boltzmann (black dotted line) and kappa-type distribution functions in arbitrary units. The latter distribution is shown for different values of the index $\kappa$. The shaded area illustrates the suprathermal tail of the kappa-type distribution function.}
\label{ab:distri}
\end{figure}

According to the linearized Vlasov theory \citep[see, e.\,g.,][for an introduction]{dav83:kin,rs:rays,tau12:nov}, magnetic field growth can be described in terms of plasma instabilities. By limiting ourselves to the simplified cases of parallel and perpendicular wave vectors with respect to a given symmetry axis, three dispersion functions \citep[see][]{tau05:co1,tau06:co2} can be derived, which describe: (i) the parallel longitudinal mode, $D_\ell$; (ii) the parallel transverse mode, $D_t$; and (iii) the perpendicular ordinary wave mode, $D_\perp$. The solution of the dispersion relations, $D_i=0$, then yields the (complex) frequency as a function of the wave vector, where the imaginary frequency part describes growth (if positive) or damping (if negative). For the velocity distribution functions introduced in subsection~\ref{tech:dist}, the dispersion relations are summarized in Appendix~\ref{app:disprel}.

\subsection{Maximum growth rate}

Here, however, we are more interested in the \emph{maximum} growth rate, which, according to $\tau\sim\Ga^{-1}$, has the shortest characteristic growth time.
By using the implicit function theorem, the maximum growth rate can be directly obtained \citep{tau11:max} together with the associated ``maximum'' wavenumber from
\be
\dd[\omega]k=-\frac{\partial D(\omega,k)/\partial k}{\partial D(\omega,k)/\partial\omega},
\ee
which, by requiring $\df\omega/\df k=0$ in order to have an extremum of the growth rate as a function of the wavenumber, can be expressed as $\partial D(\omega,k)/\partial k=0$ together with $D(\omega,k)=0$. Thus, the maximum growth rate and the associated wavenumber are obtained as
\begin{align}
&D(\omega,k)=0\;\wedge\;\partial D(\omega,k)/\partial k=0\nonumber\\
\Rightarrow\quad&\{k_{\text{max}},\,\Ga_{\text{max}}\}. \label{eq:kGmax}
\end{align}
The typical shape of the growth rate (see Fig.~\ref{ab:growthrate}) ensures that the result indeed corresponds to the maximum growth rate. While the largest unstable wavenumber, $k_{\text{end}}$, can often be determined analytically, the unstable wavenumber associated with the maximum growth rate, $k_{\text{max}}$, is available only by numerically solving Eqs.~\eqref{eq:kGmax}.

\subsection{Distribution function}\label{tech:dist}

\begin{figure}[tb]
\centering
\includegraphics[width=\linewidth]{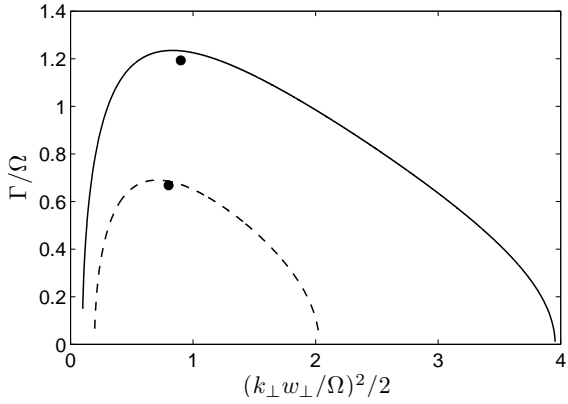}
\caption{Growth rate as obtained from the analytical solution of the dispersion relation for perpendicular wave propagation (lines) in comparison to the maximum growth rate resulting from a PIC simulation (dots). The solid and dashed lines correspond to the cases of $T\pa=T\se$ and $T\pa=2T\se$, respectively, with $(v_0/w\pa)^2=2$ in both cases \citep[cf.][]{tau06:co2,tau07:amp}.}
\label{ab:pic_perp}
\end{figure}

The required free energy is provided through anisotropies in the form of: (i) two interpenetrating (``counterstreaming'') components with (ii) different temperatures in the directions parallel and perpendicular to the streaming direction. For the initial distribution function, two different forms are assumed. The first one is a Maxwellian \citep{tau05:co1,tau06:co2},
\be\label{eq:dist_mw}
f(\f v)=C^{-1}\exp\!\left(-\frac{v\se^2}{w\se^2}\right)\sum_{j=\pm1}\exp\!\left[-\frac{\left(v\pa+jv_0\right)^2}{w\pa^2}\right],\!
\ee
where
\be\label{eq:wps}
w_{\parallel,\perp}=\sqrt{\frac{2\kB T_{\parallel,\perp}}{m}}
\ee
are the thermal velocities and where $C=2\pi^{3/2}w\se^2w\pa$ is a constant factor to ensure the normalization of $f$.

The second one is a kappa-type distribution to include suprathermal particles \citep{laz08:ka1,laz10:ka2}
\be\label{eq:dist_kap}
f(\f v)=\tilde C^{-1}\sum_{j=\pm1}\left[1+\frac{v\se^2}{\kappa\theta\se^2}+\frac{\left(v\pa+jv_0\right)^2}{\kappa\theta\pa^2}\right]^{-(\kappa+1)},
\ee
where the index $\kappa$ characterizes the fraction of suprathermal particles. The normalization factor is now given as
\begin{equation*}
\tilde C=2\pi^{3/2}\theta\se^2\theta\pa\,\frac{\kappa^{3/2}\Ga(\kappa-1/2)}{\Ga(\kappa+1)}
\end{equation*}
and where the (modified) thermal velocities are
\be
\theta_{\parallel,\perp}=\sqrt{\frac{2\kappa-3}{2\kappa}}\,w_{\parallel,\perp}.
\ee

Note that both distribution functions are limited to non-relativistic temperatures and streaming velocities. Relativistic \citep{tau05:cov,tau10:cov} and semi-relativistic \citep{zah07:sem,tau08:wei} generalizations would require the use of relativistic dispersion relations and are, therefore, considerably more difficult. Relativistic effects may be extremely important \citep{tau06:har} but only if the relevant parameters are truly relativistic \citep{usr05:cov}.

\section{Limit Estimation}\label{estim}

The maximum growth rate and the corresponding wavenumber in Eq.~\eqref{eq:kGmax} can be used to estimate the maximum turbulent magnetic field strength generated by the instability. Additionally, the associated spatial scales can be determined. Consider both in turn.

\subsection{Maximum magnetic field strength}

According to \citet{sch05:ori}, the maximum field strength can be estimated from the condition that the Larmor radius in the generated magnetic field strength be comparable to the characteristic length scale. The latter is given through the wavenumber for which the growth is maximal, which leads to
\be
\Rl\sim k_{\text{max}}^{-1}.
\ee
With $\Rl=v/\Om$, where $v=v_0$ is the streaming velocity and where $\Om=qB_{\text{max}}/(mc)$ is assumed to be the gyrofrequency that results from the emergent field, the maximum magnetic field strength can be estimated to
\be\label{eq:Bmax1}
B_{\text{max}}\sim\eta\,\frac{mc}{q}\,v_0k_{\text{max}}.
\ee
Note that the presence of a background magnetic field strength requires a considerably more complex calculation \citep[e.\,g.,][]{kat05:sat} involving the currents induced by the background magnetic field. However, for the parameters typically found in the interstellar medium, the strongest magnetic fields usually stem from the longitudinal mode as shown below, which is unaffected by the presence of a background magnetic field.

The additional factor, $\eta\sim0.01$, is due to the fact that numerical simulations typically show a somewhat reduced maximum magnetic field strength \citep{sch05:ori}. The maximum growth rate and the corresponding wavenumber, in contrast, have been reproduced with fairly good accuracy, as confirmed by Fig.~\ref{ab:pic_perp}. Both the positions and the magnitudes of the maximum growth rate are in agreement with each other \citep[see][]{tau06:co2,tau07:amp}.

For electrons, the maximum magnetic field resulting from Eq.~\eqref{eq:Bmax1} can be expressed as
\bs\label{eq:BmaxEl}
\be
\left(\frac{B_{\text{max}}}{\text{\textmu G}}\right)\approx56.856\left(\frac{v_0}{\text{km/s}}\right)\left(\frac{k_{\text{max}}}{\text{cm}^{-1}}\right),
\ee
for electrons or, alternatively, as
\be
\left(\frac{B_{\text{max}}}{\text{\textmu G}}\right)\approx1.07\!\times\!10^{-4}\left(\frac{v_0}{\text{km/s}}\right)\!\left(\frac{n_e}{\text{cm}^{-3}}\right)^{1/2}\tilde k_{\text{max}},
\ee
\es
where the usual normalization $\tilde k=ck/\omega_{\text p,e}$ is employed with $\omega_{\text p,e}=\sqrt{4\pi n_eq^2/m_e}$ the electron plasma frequency. The fact that the largest unstable wavenumber, $k_{\text{end}}$, is typically of the same order of magnitude (cf. Fig.~\ref{ab:growthrate}) as the maximum unstable wavenumber, $k_{\text{max}}$, allows one to use the first as a rough estimate, whereas the second gives the more precise result.

\begin{table}[t]
\caption{Parameter values for the streaming velocity, $v_0$, the temperature, $T$, the particle number density, $n$, and the ambient magnetic field strength, $B_0$.}
\begin{tabular}{@{}lll}
\hline
Symbol\!\! & Values & Reference \\
\hline
$v_0$ 	& 10 -- 20\,km/s 						& \citet{neh08:clo}\\
$v_0$ 	& 60 -- 150\,km/s						& \citet{aal99:map}\\
$v_0$	& $0.012\,c$ -- $0.12\,c$				& \citet{zwe82:mol}\\
$T$		& 10 -- $10^7$\, K						& \citet{kar07:fun}\\
$n$		& $10^{-1}$ -- $10^6\,\text{cm}^{-3}$\!\!	& ---\\
$B_0$	& $3\times10^{-6}$\,G					& \citet{bec96:mag}\\
\hline
\end{tabular}
\end{table}

\begin{figure}[!ht]
\centering
\includegraphics[bb=110 110 484 734,clip,width=0.98\linewidth]{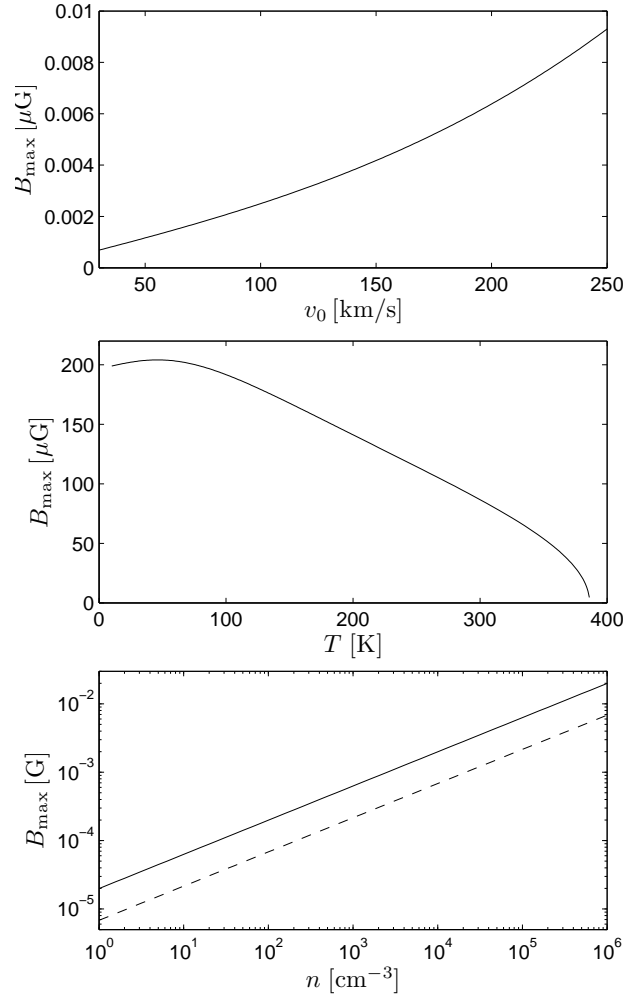}
\caption{Maximum magnetic field strength generated by the longitudinal dispersion relation as a function of different parameters. In the \emph{upper panel}, the streaming velocity, $v_0$, is varied, while other parameters are chosen as $T=10^5$\,K and $n=10^2\,\text{cm}^{-3}$. In the \emph{middle panel}, $v_0=100$\,km/s and $n=10^2\,\text{cm}^{-3}$. In the \emph{lower panel}, $v_0=100$\,km/s and $T=10$\,K (solid line) as well as $v_0=150$\,km/s and $T=10^5$\,K (dashed line).}
\label{ab:BmaxLong}
\end{figure}

\subsection{Instability scales}

From the unstable wavenumber range (cf. Fig.~\ref{ab:pic_perp}), the spatial scales of the generated magnetic field structures can be estimated to $L_{\text{min}}=2\pi/k_{\text{end}}\leqslant L\leqslant L_{\text{max}}=2\pi/k_{\text{min}}$. Even though usually $k_{\text{min}}\to0$, the maximum spatial scale, $L_{\text{max}}$ will be finite, being set by the size of the localized intense gaseous streaming region \citep{sch05:ori}. A mean scale is given by
\be
\langle L\rangle=\frac{2\pi}{k_{\text{max}}}=3.339\times10^6\left(\frac{n_e}{\text{cm}^{-3}}\right)^{-1/2}\tilde k_{\text{max}}^{\;-1}.
\ee
With $\tilde k_{\text{max}}$ being of the order unity, this corresponds to $1000\,\text{km}\lesssim\langle L\rangle\lesssim10\,\text{m}$ for $10^{-3}\,\text{cm}^{-3}\lesssim n_e\lesssim10^7\,\text{cm}^{-3}$.

\section{Parameter Study}\label{param}

In this section, the resulting maximum magnetic field strength is presented that can be expected from the saturation condition of the linear instability phase. The case of a Maxwellian distribution and the modifications introduced by a supra-thermal particle population will be discussed in turn.

\subsection{Maxwellian distribution}

\begin{figure}[tb]
\centering
\includegraphics[bb=97 233 484 610,clip,width=\linewidth]{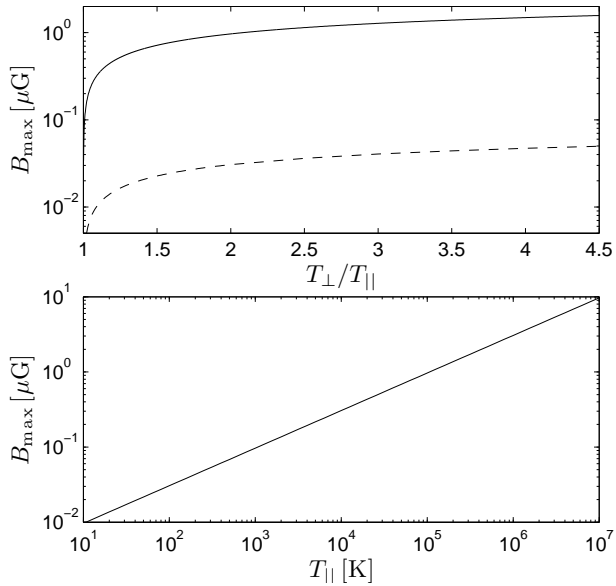}
\caption{Maximum magnetic field strength for the parallel transverse dispersion relation. The \emph{upper panel} illustrates influence of the temperature anisotropy on the maximum magnetic field strength for $T\pa=10^5$\,K (solid line) and $T\pa=10^2$\,K (dashed line). In the \emph{lower panel}, the huge influence of the thermal spread in the particle ensemble is demonstrated for $T\se/T\pa=2$. In both panels, the particle density is chosen as $n=10^2\,\text{cm}^{-3}$.}
\label{ab:BmaxTrans}
\end{figure}

In the upper and middle panels of Fig.~\ref{ab:BmaxLong}, the maximum magnetic field strength as obtained from Eqs.~\eqref{eq:BmaxEl} is shown as a function of the counterstreaming velocity, $v_0$, and as a function of the temperature entering the thermal velocity, respectively. For large temperatures and a moderate streaming velocities, the resulting maximum magnetic field strength is relatively low, as confirmed by the upper panel of Fig.~\ref{ab:BmaxLong}, where $v_0\ll w$ with $w$ the thermal velocity as defined in Eq.~\eqref{eq:wps}. In contrast, the influence of the (in this case: isotropic) temperature on the resulting field strength for the case $w\lesssim v_0$ is shown in the middle panel. This comparison confirms that, for thermal velocities exceeding the streaming velocity, the instability rate is drastically reduced as the thermal spread is no longer the dominant feature of the particle distribution.

\begin{figure}[tb]
\centering
\includegraphics[bb=97 230 487 610,clip,width=\linewidth]{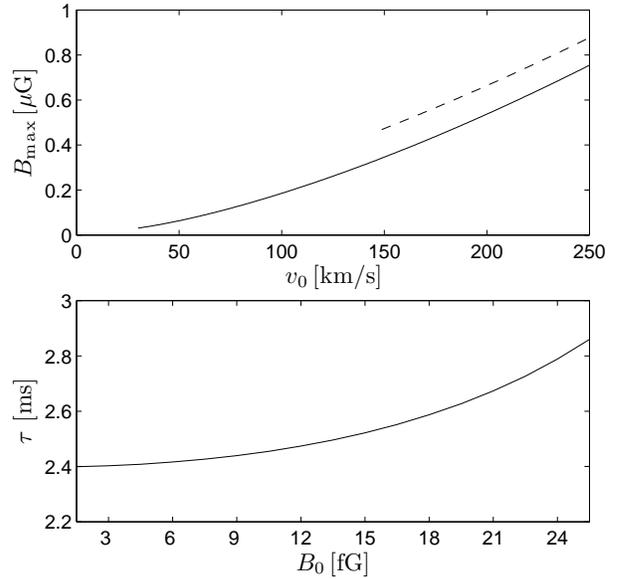}
\caption{Maximum magnetic field strength and growth time for the transverse, ordinary-mode dispersion relation. In the \emph{upper panel}, an ambient magnetic field is shown to have only a moderate influence by the comparison of the two cases with $B_0=3$\,fG (solid line) and $B_0=25.5$\,fG (dashed line). In the \emph{lower panel}, the growth time---i.\,e., the inverse growth rate---is illustrated for a variable background magnetic field, with $v_0=250$\,km/s fixed. In both panels, other parameters are chosen as $T=20$\,K and $n=10^2\,\text{cm}^{-3}$.}
\label{ab:BmaxPerp}
\end{figure}

In addition, the lower panel in Fig.~\ref{ab:BmaxLong} depicts the maximum field strength for two different temperatures as the particle density (which enters both the growth and the normalized wavenumber via the plasma frequency) is varied. Note that, for normalized variables as used in the dispersion relations (cf. Appendix~\ref{app:disprel}), Eq.~\eqref{eq:BmaxEl} is the only density dependence---at least if collisions are completely neglected as done throughout the derivation of the dispersion relations.

In the upper panel of Fig.~\ref{ab:BmaxTrans}, the effect of the temperature anisotropy on the transverse mode is shown for two different parallel temperature values. Note that, in contrast to the longitudinal two-stream mode, here a higher temperature \emph{increases} the resulting magnetic field so that, in the limit of a cold plasma, the transverse Weibel instability is suppressed. In addition, while the maximum field strength is saturated as the temperature ratio is increased, the parallel thermal velocity has a steady influence with $B_{\text{max}}\propto T\pa^{1/2}$ as confirmed in the lower panel. While the longitudinal mode is insensitive to the presence of a homogeneous background magnetic field, the transverse Weibel instability is modified in that the resulting modes are no longer aperiodic. Instead, they have a real frequency part, $\omega_r\approx\Om$ with $\Om=qB_0/(mc)$ the gyro-frequency \citep{tau08:wei}; these have been named mirror modes.

In contrast, the ordinary-mode wave---also known as the filamentation instability---is generated \emph{only} in the presence of an ambient magnetic field. However, depending on the plasma parameters, this background field must not be too strong because otherwise the instability will be suppressed \citep{tau07:amp,sto07:fil,sto08:pic}. Fig.~\ref{ab:BmaxPerp} illustrates that a stronger background magnetic field results in a stronger instability; however, there is a critical background magnetic field strength above which the instability is quickly suppressed \citep[Fig.~5 in][]{tau11:max}. In the limit of a cold plasma, the critical magnetic field strength is given by \citep{sto08:pic}
\bs\label{eq:Bcrit}
\begin{align}
B_{\text{crit}}&=\omega_{\text p}\,\frac{m}{qc}\,\frac{v_0}{\sqrt\gamma}\\
&\approx3.569\times10^{-13}\left(\frac{v_0}{\text{km/s}}\right)\left(\frac{n}{\text{cm}^{-3}}\right)^{1/2}.
\end{align}
\es
In contrast to the Weibel instability, both the growth rate and the maximum unstable wavenumber are decreased for a warm plasma; accordingly, $B_{\text{crit}}$ as given by Eq.~\eqref{eq:Bcrit} represents an upper level to the critical field strength. Thus, for non-relativistic streaming velocities and moderate densities, the instability will almost always be suppressed. Apart from the suppression, the background magnetic field has only a moderate influence both on the resulting maximum turbulent field strength and on the instability growth time, as confirmed in both panels of Fig.~\ref{ab:BmaxPerp}.

\begin{figure}[tb]
\centering
\includegraphics[bb=92 222 485 610,clip,width=\linewidth]{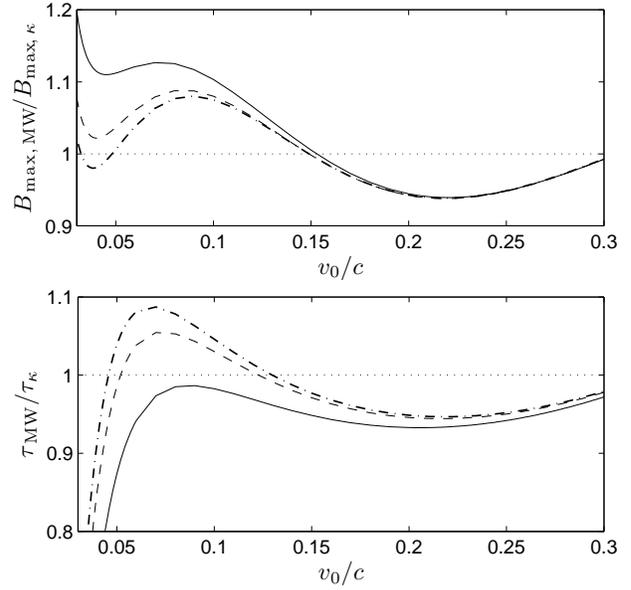}
\caption{Modifications of the longitudinal mode introduced by the use of a kappa distribution function with $\kappa=2$ (solid lines), $\kappa=3$ (dashed lines), and $\kappa=4$ (dotdashed lines). Shown are the maximum magnetic field strengths \emph{(upper panel)} and the instability growth times \emph{(lower panel)} in relation to the Maxwellian case.}
\label{ab:BmaxTauKappa}
\end{figure}

\subsection{Suprathermal distribution}

The effect of a suprathermal particle population is illustrated in Fig.~\ref{ab:BmaxTauKappa}. For the longitudinal mode, the upper panel shows the ratio of the respective maximum magnetic field strengths as the counterstreaming velocity is varied. It is confirmed that, for larger values of the spectral index, $\kappa$, the difference becomes less pronounced.

In the lower panel of Fig.~\ref{ab:BmaxTauKappa}, the characteristic instability growth times are compared, which are simply given by $\tau=\Ga^{-1}$ with $\Ga$ the growth rate. The observation is that, depending on the precise choice of the instability parameters \emph{and} the spectral index of the supra-thermal tail, the instability can growth faster or slower compared to the Maxwellian case.

From the detailed investigation of the kappa-type distribution \citep{laz08:ka1,laz10:ka2}, it is known that, for parallel wave propagation, the Maxwellian distribution provides an upper limit to the growth rate---and, accordingly, a lower limit to the instability growth time; for perpendicular wave propagation, the situation is reversed so that the growth rate exceeds that for the Maxwellian. However, as shown in Fig.~\ref{ab:BmaxTauKappa}, the correction factors are usually close to unity if the streaming velocity is not too small.

\section{Summary and Conclusion}\label{summ}

In this paper, three linear plasma instabilities have been investigated, which are: (i) the longitudinal two-stream instability, (ii) the classic Weibel instability, and (iii) the perpendicular filamentation instability with the latter two being transverse modes. In contrast to previous investigations that relied on normalized parameters and determined the instability growth rate (typically in relation to the plasma frequency) as a function of the wavenumber, here the maximum turbulent magnetic field strength, $B_{\text{max}}$ has been investigated that can be generated by these unstable modes. Whereas an exact determination of $B_{\text{max}}$ requires knowledge of the non-linear behavior of the instability, a simple estimate has been used that involves only the wavenumber associated with the maximum growth rate.

For parameter values that are typical for environments such as molecular clouds or the diffusive interstellar medium, the following main results have been found:
\begin{itemize}
\item for the longitudinal mode (two-stream instability), the temperature has an overarching influence, which can be understood when bearing in mind that the \emph{ratio} of oriented streaming and random thermal motion dictates the resulting instability rate. For low temperatures, therefore, magnetic field generation is considerably more efficient;
\item the transverse electromagnetic mode (classic Weibel instability), in contrast, is more efficient for high plasma temperatures with the perpendicular temperature being significantly higher than the parallel temperature;
\item the perpendicular ordinary-wave mode (filamentation instability) is most efficient for cold plasmas. Finite temperatures decrease the resulting instability rate but tend to stabilize the mechanism provided that the parallel temperature exceeds the perperpendicular temperature;
\item for all modes, the maximum field strength scales with the square root of the particle density; however, in that regard it has to be noted that collisional effects have been neglected throughout.
\item while an ambient magnetic field leaves the longitudinal mode unaffected, the other modes are modified in that: (i) the transverse mode has now an oscillation frequency of the order of the gyrofrequency; and (ii) the perpendicular mode is, for the parameters considered here, suppressed even for a magnetic field as low as $\lesssim0.1$\,nG.
\end{itemize}

In any case, it has been found that, depending (mostly) on the particle density, turbulent magnetic fields of the order of micro Gauss can be generated. Such is in agreement with the usual assumption of the turbulent magnetic field strength having the same order of magnitude as the background magnetic field \citep[e.\,g.,][]{sof86:mag}. Even though the present-day interstellar medium is generally magnetized, the results obtained here are relevant if plasmas counterstream along magnetic field lines and/or in circumstances when the thermal plasma energy density exceeds the magnetic field energy density (i.\,e., high plasma~$\beta$).

An additional uncertainty is owed to the fact that only the cases have been investigated with unstable modes oriented parallel and perpendicular to a given symmetry axis. It has been known that the fastest growing mode usually has an oblique axis of wave propagation \citep{die06:mix}. In that case, however, mode-coupling effects have to be taken into account \citep{tau07:har,tau12:3x3}. Furthermore, the presence of multiple particle species such as electrons, positrons, and ions introduce additional effects in the resulting growth rate \citep{tau08:ba1}. These effects, which will modify the maximum magnetic field strengths presented here, will therefore be incorporated in future work.

\acknowledgments
J.\,T. thanks D. Breit\-schwerdt for the supervision of her thesis.

\appendix
\section{Dispersion Relations}\label{app:disprel}

For a any given distribution function with a specified anisotropy pattern, the dispersion relations $D_\ell$, $D_t$, and $D\se$ can be evaluated. The resulting equations relating $\omega\in\C$ and $k\pa$ or $k\se\in\R$ are usually non-linear and often transcendental. In most cases, a numerical solution is required, even though a series expansions can often lead to reasonable approximative solutions.

It should be mentioned that there are investigations without specifying a distribution function \citep[e.\,g.,][]{usr06:wei,tau12:3x3}, which has shed light on the general behavior of the instability. A comparison of the instability for various distribution functions \citep{usr08:wei} has shown that the mechanism is indeed robust and does not strongly depend on the precise form of the distribution, as long as the anisotropy clearly dominates over the thermal spread of the particle ensemble.

Furthermore, note that all dispersion relations are valid in the non-relativistic regime only, i.\,e., for counterstreaming and thermal velocities small compared to the speed of light. For a discussion of relativistic effects see, e.\,g., \citet{usr05:cov,tau05:cov,tau12:3x3}.

\subsection{Maxwellian distribution}

For the Maxwellian distribution from Eq.~\eqref{eq:dist_mw}, the temperature- and $\kappa$-dependent parameters $\theta\pa$ and $\theta\se$ now play the role of the thermal velocities in the sense that, by calculating the first moment of the distribution, the appropriate $\theta$ is obtained.

For a plasma consisting of multiple particle species (denoted with the index $a$), the longitudinal dispersion relation, $D_\ell$, reads \citep{tau05:co1}
\be
D_\ell=k^2-\frac{1}{2}\sum_a\left(\frac{\omega_{\text p,a}}{w\pa}\right)^2\left[Z'\!\left(\frac{\omega-v_0k}{kw\pa}\right)+Z'\!\left(\frac{\omega+v_0k}{kw\pa}\right)\right]=0,
\ee
where $Z'$ is the derivative of the plasma dispersion function,
\be
Z(x)=\frac{1}{\sqrt\pi}\uint\df t\;\frac{e^{-t^2}}{t-x}=i\sqrt\pi\,e^{-x^2}\left[1+\erf{ix}\right],
\ee
where the first form is valid for $\Im(x)>0$ only and where erf denotes the error function.

For the transverse dispersion relation, there are two versions if a background magnetic field, \bo, is present---the left-handed and right-handed modes---which can be expressed as \citep{tau05:co1}
\begin{align}
D_t^\pm&=\omega^2-c^2k^2-\sum_a\omega_{\text p,a}^2\mp\frac{1}{2}\sum_a\frac{\omega_{\text p,a}^2}{w\pa}\frac{\Om}{k}\left[Z\!\left(\frac{\omega-kv_0\pm\Om}{kw\pa}\right)+Z\!\left(\frac{\omega+kv_0\pm\Om}{kw\pa}\right)\right]\nonumber\\
&-\frac{1}{4}\sum_a\omega_{\text p,a}^2\left(\frac{w\se}{w\pa}\right)^2\left[Z'\!\left(\frac{\omega-kv_0\pm\Om}{kw\pa}\right)+Z'\!\left(\frac{\omega+kv_0\pm\Om}{kw\pa}\right)\right].
\end{align}
Note that, due to the linear factor $\Om$ in the terms containing $Z(\dots)$, the dispersion relation is greatly simplified for an unmagnetized plasma, i.\,e., where $B_0=0$.

For perpendicular wave propagation, the dispersion relation for the ordinary-wave mode reads \citep{tau06:co2}
\be\label{eq:Dse_MW}
D\se=\omega^2-c^2k^2+\sum_a\omega_{\text p,a}^2+\sum_a\omega_{\text p,a}^2\,\frac{w\pa^2+2v_0^2}{w\se^2}\left[1-\Ff{\frac{1}{2},1;1+\frac{\omega}{\Om},1-\frac{\omega}{\Om};-\frac{k^2w\se^2}{\Om^2}}\right],
\ee
where $\leftidx{_2}{F}{_2}(a,b;c,d;z)$ is the generalized hypergeometric function.

\subsection{Suprathermal distribution}

For the kappa-type distribution function that includes particles forming a so-called supra-thermal tail, the longitudinal dispersion relation reads \citep{laz08:ka1}
\be
D_\ell=k^2+\sum_a\frac{\omega_{\text p,a}^2}{\theta\pa^2}\left[2-\frac{1}{\kappa}+\frac{\omega-kv_0}{k\theta\pa}\,Z_\kappa\!\left(\frac{\omega-kv_0}{k\theta\pa}\right)+\frac{\omega+kv_0}{2k\theta\pa}\,Z_\kappa\!\left(\frac{\omega+kv_0}{2k\theta\pa}\right)\right],
\ee
where one must take care not to confuse $\kappa$ (the power-law index in the distribution function) with $k$ (the wavenumber). The modified plasma dispersion function \citep{sum91:mod} is given through
\be\label{eq:Zk}
Z_\kappa(x)=\frac{1}{\sqrt{\pi\kappa}}\,\frac{\Ga(\kappa)}{\Ga(\kappa-1/2)}\uint\df t\;\frac{\left(1+x^2/\kappa\right)^{-(\kappa+1)}}{t-x},
\ee
where $\Ga(z)$ is the Gamma function. Again, Eq.~\eqref{eq:Zk} is valid for $\Im(x)>0$ only.

For the transverse dispersion relation, the form that has been derived by \citet{laz08:ka1} is given as
\begin{align}
D_t^\pm&=\omega^2-c^2k^2-\sum_a\omega_{\text p,a}^2\mp\frac{1}{2}\sum_a\frac{\omega_{\text p,a}^2}{\theta\pa}\frac{\Om}{k}\left[\tilde Z_\kappa\!\left(\frac{\omega-kv_0\pm\Om}{k\theta\pa}\right)+\tilde Z_\kappa\!\left(\frac{\omega+kv_0\pm\Om}{k\theta\pa}\right)\right]\nonumber\\
&+\frac{1}{2}\sum_a\omega_{\text p,a}^2\left(\frac{\theta\se}{\theta\pa}\right)^2\left[2+\frac{\omega-kv_0\pm\Om}{k\theta\pa}\,\tilde Z_\kappa\!\left(\frac{\omega-kv_0\pm\Om}{k\theta\pa}\right)+\frac{\omega+kv_0\pm\Om}{k\theta\pa}\,\tilde Z_\kappa\!\left(\frac{\omega+kv_0\pm\Om}{k\theta\pa}\right)\right],
\end{align}
where, for the plasma dispersion function, a new form has been introduced as
\bs
\be
\tilde Z_\kappa(x)=\frac{1}{\sqrt{\pi\kappa}}\,\frac{\Ga(\kappa)}{\Ga(\kappa-1/2)}\uint\df t\;\frac{\left(1+x^2/\kappa\right)^{-\kappa}}{t-x},
\ee
which is related to $Z_\kappa(x)$ in Eq.~\eqref{eq:Zk} as
\be
\tilde Z_\kappa(x)=\left(1+\frac{x^2}{\kappa}\right)Z_\kappa(x)+\frac{x}{\kappa}\left(1-\frac{1}{2\kappa}\right).
\ee
\es

In general, the ordinary-wave mode for perpendicular wave propagation would involve a rather tedious integral. Therefore, \citet{laz10:ka2} used the large-wavelength limit, in which case the dispersion relation can be written in simplified form as
\be
D\se(k\Rl\ll1)\approx\omega^2-c^2k^2-\sum_a\omega_{\text p,a}^2-k^2\sum_a\frac{\omega_{\text p,a}^2v_0^2}{\omega^2-\Om^2}\left[1+\left(\frac{w\se}{v_0}\right)^2\right],
\ee
which agrees with the corresponding expansion of the dispersion relation for the case of a Maxwellian distribution function, Eq.~\eqref{eq:Dse_MW}. In the opposite limit of small wavelengths, the dispersion relation reads
\be
D\se(k\Rl\gg1)\approx\omega^2-c^2k^2-\sum_a\omega_{\text p,a}^2+\sum_a\omega_{\text p,c}^2\left(\frac{\theta\pa}{\theta\se}\right)^2\left[1+\left(2-\frac{1}{\kappa}\right)\frac{v_0^2}{\theta\pa^2}\right].
\ee
A discussion of the applicability and numerical solutions connecting the two limiting cases has been given by \citet{laz10:ka2}.


\begin{thebibliography}{58}
\ifx \bisbn   \undefined \def \bisbn  #1{ISBN #1}\fi
\ifx \binits  \undefined \def \binits#1{#1} \fi
\ifx \bauthor  \undefined \def \bauthor#1{#1} \fi
\ifx \batitle  \undefined \def \batitle#1{#1} \fi
\ifx \bjtitle  \undefined \def \bjtitle#1{#1}\fi
\ifx \bvolume  \undefined \def \bvolume#1{\textbf{#1}}\fi
\ifx \byear  \undefined \def \byear#1{#1} \fi
\ifx \bissue  \undefined \def \bissue#1{#1} \fi
\ifx \bfpage  \undefined \def \bfpage#1{#1} \fi
\ifx \blpage  \undefined \def \blpage #1{#1} \fi
\ifx \burl  \undefined \def \burl#1{\textsf{#1}} \fi
\ifx \doiurl  \undefined \def \doiurl#1{\textsf{#1}} \fi
\ifx \betal  \undefined \def \betal{\textit{et al.}} \fi
\ifx \binstitute  \undefined \def \binstitute#1{#1} \fi
\ifx \binstitutionaled  \undefined \def \binstitutionaled#1{#1} \fi
\ifx \bctitle  \undefined \def \bctitle#1{#1} \fi
\ifx \beditor  \undefined \def \beditor#1{#1} \fi
\ifx \bpublisher  \undefined \def \bpublisher#1{#1} \fi
\ifx \bbtitle  \undefined \def \bbtitle#1{#1} \fi
\ifx \bedition  \undefined \def \bedition#1{#1} \fi
\ifx \bseriesno  \undefined \def \bseriesno#1{#1} \fi
\ifx \blocation  \undefined \def \blocation#1{#1} \fi
\ifx \bsertitle  \undefined \def \bsertitle#1{#1} \fi
\ifx \bsnm \undefined \def \bsnm#1{#1} \fi
\ifx \bsuffix \undefined \def \bsuffix#1{#1} \fi
\ifx \bparticle \undefined \def \bparticle#1{#1} \fi
\ifx \barticle \undefined \def \barticle#1{#1} \fi
\ifx \bconfdate \undefined \def \bconfdate #1{#1} \fi
\ifx \botherref \undefined \def \botherref #1{#1} \fi
\ifx \url \undefined \def \url#1{\textsf{#1}} \fi
\ifx \bchapter \undefined \def \bchapter#1{#1} \fi
\ifx \bbook \undefined \def \bbook#1{#1} \fi
\ifx \bcomment \undefined \def \bcomment#1{#1} \fi
\ifx \oauthor \undefined \def \oauthor#1{#1} \fi
\ifx \citeauthoryear \undefined \def \citeauthoryear#1{#1} \fi
\ifx \endbibitem  \undefined \def \endbibitem {}\fi
\ifx \bconflocation  \undefined \def \bconflocation#1{#1} \fi
\ifx \arxivurl  \undefined \def \arxivurl#1{\textsf{#1}} \fi

\bibitem[\protect\citeauthoryear{Aalto et~al.}{1999}]{aal99:map}
\begin{barticle}
\bauthor{\bsnm{Aalto}, \binits{S.}},
\bauthor{\bsnm{H\"uttemeister}, \binits{S.}},
\bauthor{\bsnm{Scoville}, \binits{N.Z.}},
\bauthor{\bsnm{Thaddeus}, \binits{P.}}:
\bjtitle{Astrophys. J.}
\bvolume{522},
\bfpage{165}
(\byear{1999})
\end{barticle}
\endbibitem

\bibitem[\protect\citeauthoryear{Achterberg and Wiersma}{2007}]{ach07:we1}
\begin{barticle}
\bauthor{\bsnm{Achterberg}, \binits{A.}},
\bauthor{\bsnm{Wiersma}, \binits{J.}}:
\bjtitle{Astron. Astrophys.}
\bvolume{475},
\bfpage{1}
(\byear{2007})
\end{barticle}
\endbibitem

\bibitem[\protect\citeauthoryear{Achterberg et~al.}{2007}]{ach07:we2}
\begin{barticle}
\bauthor{\bsnm{Achterberg}, \binits{A.}},
\bauthor{\bsnm{Wiersma}, \binits{J.}},
\bauthor{\bsnm{Norman}, \binits{C.A.}}:
\bjtitle{Astron. Astrophys.}
\bvolume{475},
\bfpage{19}
(\byear{2007})
\end{barticle}
\endbibitem

\bibitem[\protect\citeauthoryear{Beck et~al.}{1996}]{bec96:mag}
\begin{barticle}
\bauthor{\bsnm{Beck}, \binits{R.}},
\bauthor{\bsnm{Brandenburg}, \binits{A.}},
\bauthor{\bsnm{Moss}, \binits{D.}},
\bauthor{\bsnm{Shukurov}, \binits{A.}},
\bauthor{\bsnm{Sokoloff}, \binits{D.}}:
\bjtitle{Annu. Rev. Astron. Astrophys.}
\bvolume{34},
\bfpage{155}
(\byear{1996})
\end{barticle}
\endbibitem

\bibitem[\protect\citeauthoryear{Bernet et~al.}{2008}]{ber08:gal}
\begin{barticle}
\bauthor{\bsnm{Bernet}, \binits{M.L.}},
\bauthor{\bsnm{Miniati}, \binits{F.}},
\bauthor{\bsnm{Lilly}, \binits{S.J.}},
\bauthor{\bsnm{Kronberg}, \binits{P.P.}},
\bauthor{\bsnm{Dessauges–Zavadsky}, \binits{M.}}:
\bjtitle{Nature}
\bvolume{454},
\bfpage{302}
(\byear{2008})
\end{barticle}
\endbibitem

\bibitem[\protect\citeauthoryear{Brandenburg and Subramanian}{2005}]{bra05:dyn}
\begin{barticle}
\bauthor{\bsnm{Brandenburg}, \binits{A.}},
\bauthor{\bsnm{Subramanian}, \binits{K.}}:
\bjtitle{Phys. Rep.}
\bvolume{417},
\bfpage{1}
(\byear{2005})
\end{barticle}
\endbibitem

\bibitem[\protect\citeauthoryear{Bret et~al.}{2006}]{bre06:obl}
\begin{barticle}
\bauthor{\bsnm{Bret}, \binits{A.}},
\bauthor{\bsnm{Dieckmann}, \binits{M.E.}},
\bauthor{\bsnm{Deutsch}, \binits{C.}}:
\bjtitle{Phys. Plasmas}
\bvolume{13},
\bfpage{082109}
(\byear{2006})
\end{barticle}
\endbibitem

\bibitem[\protect\citeauthoryear{Clemmow and Dougherty}{1969}]{cle69:ele}
\begin{bbook}
\bauthor{\bsnm{Clemmow}, \binits{P.C.}},
\bauthor{\bsnm{Dougherty}, \binits{J.P.}}:
\bbtitle{{Electrodynamics of Particles and Plasmas}}.
\bpublisher{Addison-Wesley},
\blocation{Reading, MA}
(\byear{1969})
\end{bbook}
\endbibitem

\bibitem[\protect\citeauthoryear{Davidson}{1983}]{dav83:kin}
\begin{bchapter}
\bauthor{\bsnm{Davidson}, \binits{R.C.}}:
In: \beditor{\bsnm{Rosenbluth}, \binits{M.N.}},
\beditor{\bsnm{Sagdeev}, \binits{R.Z.}} (eds.)
\bbtitle{{Handbook of Plasma Physics}}
vol. \bseriesno{1},
p. \bfpage{519}.
\bpublisher{North-Holland},
\blocation{Amsterdam}
(\byear{1983})
\end{bchapter}
\endbibitem

\bibitem[\protect\citeauthoryear{Dieckmann et~al.}{2006}]{die06:mix}
\begin{barticle}
\bauthor{\bsnm{Dieckmann}, \binits{M.E.}},
\bauthor{\bsnm{Frederiksen}, \binits{J.T.}},
\bauthor{\bsnm{Bret}, \binits{A.}},
\bauthor{\bsnm{Shukla}, \binits{P.K.}}:
\bjtitle{Phys. Plasmas}
\bvolume{13},
\bfpage{112110}
(\byear{2006})
\end{barticle}
\endbibitem

\bibitem[\protect\citeauthoryear{Durrer and Neronov}{2013}]{dur13:cos}
\begin{botherref}
\oauthor{\bsnm{Durrer}, \binits{R.}},
\oauthor{\bsnm{Neronov}, \binits{A.}}:
2013,
{Cosmological Magnetic Fields: Their Generation, Evolution and Observation}.
{to appear in Space Sci. Rev.}
\end{botherref}
\endbibitem

\bibitem[\protect\citeauthoryear{Fried}{1959}]{fri59:wei}
\begin{barticle}
\bauthor{\bsnm{Fried}, \binits{B.D.}}:
\bjtitle{Phys. Fluids}
\bvolume{2},
\bfpage{337}
(\byear{1959})
\end{barticle}
\endbibitem

\bibitem[\protect\citeauthoryear{Gremillet et~al.}{2007}]{gre07:bpi}
\begin{barticle}
\bauthor{\bsnm{Gremillet}, \binits{L.}},
\bauthor{\bsnm{B\'enisti}, \binits{D.}},
\bauthor{\bsnm{Lefebvre}, \binits{E.}},
\bauthor{\bsnm{Bret}, \binits{A.}}:
\bjtitle{Phys. Plasmas}
\bvolume{14},
\bfpage{040704}
(\byear{2007})
\end{barticle}
\endbibitem

\bibitem[\protect\citeauthoryear{Heiles and Crutcher}{2005}]{hei05:mol}
\begin{bchapter}
\bauthor{\bsnm{Heiles}, \binits{C.}},
\bauthor{\bsnm{Crutcher}, \binits{R.}}:
\bctitle{{Magnetic fields in Diffuse H\textsc{I} and Molecular Clouds}}.
In: \beditor{\bsnm{Wielebinski}, \binits{R.}},
\beditor{\bsnm{Beck}, \binits{R.}} (eds.)
\bbtitle{{Cosmic Magnetic Fields}}
vol. \bseriesno{664},
p. \bfpage{137}.
\bpublisher{Springer},
\blocation{Berlin}
(\byear{2005})
\end{bchapter}
\endbibitem

\bibitem[\protect\citeauthoryear{Inoue and Inutsuka}{2012}]{ino12:mol}
\begin{barticle}
\bauthor{\bsnm{Inoue}, \binits{T.}},
\bauthor{\bsnm{Inutsuka}, \binits{S.-I.}}:
\bjtitle{Astrophys. J.}
\bvolume{759},
\bfpage{35}
(\byear{2012})
\end{barticle}
\endbibitem

\bibitem[\protect\citeauthoryear{Karttunen et~al.}{2007}]{kar07:fun}
\begin{bbook}
\bauthor{\bsnm{Karttunen}, \binits{H.}},
\bauthor{\bsnm{Kr\"oger}, \binits{P.}},
\bauthor{\bsnm{Oja}, \binits{H.}},
\bauthor{\bsnm{Poutanen}, \binits{M.}},
\bauthor{\bsnm{Donner}, \binits{K.J.}}:
\bbtitle{{Fundamental Astronomy}}.
\bpublisher{Springer},
\blocation{Berlin}
(\byear{2007})
\end{bbook}
\endbibitem

\bibitem[\protect\citeauthoryear{Kato}{2005}]{kat05:sat}
\begin{barticle}
\bauthor{\bsnm{Kato}, \binits{T.N.}}:
\bjtitle{Phys. Plasmas}
\bvolume{12},
\bfpage{080705}
(\byear{2005})
\end{barticle}
\endbibitem

\bibitem[\protect\citeauthoryear{Kulsrud}{2010}]{kul10:gal}
\begin{barticle}
\bauthor{\bsnm{Kulsrud}, \binits{R.M.}}:
\bjtitle{Astron. Nachr.}
\bvolume{331},
\bfpage{22}
(\byear{2010})
\end{barticle}
\endbibitem

\bibitem[\protect\citeauthoryear{Lazar et~al.}{2012}]{laz12:sp3}
\begin{barticle}
\bauthor{\bsnm{Lazar}, \binits{M.}},
\bauthor{\bsnm{Yoon}, \binits{P.H.}},
\bauthor{\bsnm{Schlickeiser}, \binits{R.}}:
\bjtitle{Phys. Plasmas}
\bvolume{19},
\bfpage{122108}
(\byear{2012})
\end{barticle}
\endbibitem

\bibitem[\protect\citeauthoryear{Lazar et~al.}{2008}]{laz08:ka1}
\begin{barticle}
\bauthor{\bsnm{Lazar}, \binits{M.}},
\bauthor{\bsnm{Schlickeiser}, \binits{R.}},
\bauthor{\bsnm{Poedts}, \binits{S.}},
\bauthor{\bsnm{Tautz}, \binits{R.C.}}:
\bjtitle{Mon. Not. Royal Astron. Soc.}
\bvolume{390},
\bfpage{168}
(\byear{2008})
\end{barticle}
\endbibitem

\bibitem[\protect\citeauthoryear{Lazar et~al.}{2010}]{laz10:ka2}
\begin{barticle}
\bauthor{\bsnm{Lazar}, \binits{M.}},
\bauthor{\bsnm{Tautz}, \binits{R.C.}},
\bauthor{\bsnm{Schlickeiser}, \binits{R.}},
\bauthor{\bsnm{Poedts}, \binits{S.}}:
\bjtitle{Mon. Not. Royal Astron. Soc.}
\bvolume{401},
\bfpage{362}
(\byear{2010})
\end{barticle}
\endbibitem

\bibitem[\protect\citeauthoryear{Mantare and Cole}{2012}]{man12:dyn}
\begin{barticle}
\bauthor{\bsnm{Mantare}, \binits{M.J.}},
\bauthor{\bsnm{Cole}, \binits{E.}}:
\bjtitle{Astrophys. J.}
\bvolume{753},
\bfpage{32}
(\byear{2012})
\end{barticle}
\endbibitem

\bibitem[\protect\citeauthoryear{Murphy}{2009}]{mur09:ska}
\begin{barticle}
\bauthor{\bsnm{Murphy}, \binits{E.J.}}:
\bjtitle{Astrophys. J.}
\bvolume{706},
\bfpage{482}
(\byear{2009})
\end{barticle}
\endbibitem

\bibitem[\protect\citeauthoryear{Nehm\'e et~al.}{2008}]{neh08:clo}
\begin{barticle}
\bauthor{\bsnm{Nehm\'e}, \binits{C.}},
\bauthor{\bsnm{Gry}, \binits{C.}},
\bauthor{\bsnm{Boulanger}, \binits{F.}},
\bauthor{\bsnm{Le~Bourlot}, \binits{J.}},
\bauthor{\bparticle{Pineau~des} \bsnm{For\^ets}, \binits{G.}},
\bauthor{\bsnm{Falgarone}, \binits{E.}}:
\bjtitle{Astron. Astrophys.}
\bvolume{483},
\bfpage{471}
(\byear{2008})
\end{barticle}
\endbibitem

\bibitem[\protect\citeauthoryear{Niemiec et~al.}{2010}]{nie10:bea}
\begin{barticle}
\bauthor{\bsnm{Niemiec}, \binits{J.}},
\bauthor{\bsnm{Pohl}, \binits{M.}},
\bauthor{\bsnm{Bret}, \binits{A.}},
\bauthor{\bsnm{Stroman}, \binits{T.}}:
\bjtitle{Astrophys. J.}
\bvolume{709},
\bfpage{1148}
(\byear{2010})
\end{barticle}
\endbibitem

\bibitem[\protect\citeauthoryear{Reville et~al.}{2008}]{rev08:tra}
\begin{barticle}
\bauthor{\bsnm{Reville}, \binits{B.}},
\bauthor{\bsnm{O'Sullivan}, \binits{S.}},
\bauthor{\bsnm{Duffy}, \binits{P.}},
\bauthor{\bsnm{Kirk}, \binits{J.G.}}:
\bjtitle{Mon. Not. R. Astron. Soc.}
\bvolume{386},
\bfpage{509}
(\byear{2008})
\end{barticle}
\endbibitem

\bibitem[\protect\citeauthoryear{Ryu et~al.}{2012}]{ryu12:lss}
\begin{barticle}
\bauthor{\bsnm{Ryu}, \binits{D.}},
\bauthor{\bsnm{Schleicher}, \binits{D.R.G.}},
\bauthor{\bsnm{Treumann}, \binits{R.A.}},
\bauthor{\bsnm{Tsagas}, \binits{C.G.}},
\bauthor{\bsnm{Widrow}, \binits{L.M.}}:
\bjtitle{Space Sci. Rev.}
\bvolume{166},
\bfpage{1}
(\byear{2012})
\end{barticle}
\endbibitem

\bibitem[\protect\citeauthoryear{Schaefer-Rolffs and Lerche}{2006}]{usr06:wei}
\begin{barticle}
\bauthor{\bsnm{Schaefer-Rolffs}, \binits{U.}},
\bauthor{\bsnm{Lerche}, \binits{I.}}:
\bjtitle{Phys. Plasmas}
\bvolume{13},
\bfpage{012107}
(\byear{2006})
\end{barticle}
\endbibitem

\bibitem[\protect\citeauthoryear{Schaefer-Rolffs and
  Schlickeiser}{2005}]{usr05:cov}
\begin{barticle}
\bauthor{\bsnm{Schaefer-Rolffs}, \binits{U.}},
\bauthor{\bsnm{Schlickeiser}, \binits{R.}}:
\bjtitle{Phys. Plasmas}
\bvolume{12},
\bfpage{022104}
(\byear{2005})
\end{barticle}
\endbibitem

\bibitem[\protect\citeauthoryear{Schaefer-Rolffs and Tautz}{2008}]{usr08:wei}
\begin{barticle}
\bauthor{\bsnm{Schaefer-Rolffs}, \binits{U.}},
\bauthor{\bsnm{Tautz}, \binits{R.C.}}:
\bjtitle{Phys. Plasmas}
\bvolume{15},
\bfpage{062105}
(\byear{2008})
\end{barticle}
\endbibitem

\bibitem[\protect\citeauthoryear{Schlickeiser}{2002}]{rs:rays}
\begin{bbook}
\bauthor{\bsnm{Schlickeiser}, \binits{R.}}:
\bbtitle{{Cosmic Ray Astrophysics}}.
\bpublisher{Springer},
\blocation{Berlin}
(\byear{2002})
\end{bbook}
\endbibitem

\bibitem[\protect\citeauthoryear{Schlickeiser}{2005}]{sch05:ori}
\begin{barticle}
\bauthor{\bsnm{Schlickeiser}, \binits{R.}}:
\bjtitle{Plasma Phys. Contr. Fusion}
\bvolume{47},
\bfpage{205}
(\byear{2005})
\end{barticle}
\endbibitem

\bibitem[\protect\citeauthoryear{Schlickeiser}{2012}]{sch12:equ}
\begin{barticle}
\bauthor{\bsnm{Schlickeiser}, \binits{R.}}:
\bjtitle{Phys. Rev. Lett.}
\bvolume{109},
\bfpage{261101}
(\byear{2012})
\end{barticle}
\endbibitem

\bibitem[\protect\citeauthoryear{Schober et~al.}{2012}]{sch12:amp}
\begin{barticle}
\bauthor{\bsnm{Schober}, \binits{J.}},
\bauthor{\bsnm{Schleicher}, \binits{D.}},
\bauthor{\bsnm{Federrath}, \binits{C.}},
\bauthor{\bsnm{Klessen}, \binits{R.}},
\bauthor{\bsnm{Banerjee}, \binits{R.}}:
\bjtitle{Phys. Rev.~E}
\bvolume{85},
\bfpage{026303}
(\byear{2012})
\end{barticle}
\endbibitem

\bibitem[\protect\citeauthoryear{Sofue et~al.}{1986}]{sof86:mag}
\begin{barticle}
\bauthor{\bsnm{Sofue}, \binits{Y.}},
\bauthor{\bsnm{Fujimoto}, \binits{M.}},
\bauthor{\bsnm{Wielebinski}, \binits{R.}}:
\bjtitle{Annu. Rev. Astron. Astrophys.}
\bvolume{24},
\bfpage{459}
(\byear{1986})
\end{barticle}
\endbibitem

\bibitem[\protect\citeauthoryear{Stockem et~al.}{2008}]{sto08:pic}
\begin{barticle}
\bauthor{\bsnm{Stockem}, \binits{A.}},
\bauthor{\bsnm{Dieckmann}, \binits{M.E.}},
\bauthor{\bsnm{Schlickeiser}, \binits{R.}}:
\bjtitle{Plasma Phys. Contr. Fusion}
\bvolume{50},
\bfpage{025002}
(\byear{2008})
\end{barticle}
\endbibitem

\bibitem[\protect\citeauthoryear{Stockem et~al.}{2007}]{sto07:fil}
\begin{barticle}
\bauthor{\bsnm{Stockem}, \binits{A.}},
\bauthor{\bsnm{Lerche}, \binits{I.}},
\bauthor{\bsnm{Schlickeiser}, \binits{R.}}:
\bjtitle{Astrophys. J.}
\bvolume{659},
\bfpage{419}
(\byear{2007})
\end{barticle}
\endbibitem

\bibitem[\protect\citeauthoryear{Summers and Thorne}{1991}]{sum91:mod}
\begin{barticle}
\bauthor{\bsnm{Summers}, \binits{D.}},
\bauthor{\bsnm{Thorne}, \binits{R.M.}}:
\bjtitle{Phys. Fluids~B}
\bvolume{3},
\bfpage{1835}
(\byear{1991})
\end{barticle}
\endbibitem

\bibitem[\protect\citeauthoryear{Tautz}{2010}]{tau10:cov}
\begin{barticle}
\bauthor{\bsnm{Tautz}, \binits{R.C.}}:
\bjtitle{Astrophys. Space Sci.}
\bvolume{330},
\bfpage{69}
(\byear{2010})
\end{barticle}
\endbibitem

\bibitem[\protect\citeauthoryear{Tautz}{2011}]{tau11:max}
\begin{barticle}
\bauthor{\bsnm{Tautz}, \binits{R.C.}}:
\bjtitle{Phys. Plasmas}
\bvolume{18},
\bfpage{012101}
(\byear{2011})
\end{barticle}
\endbibitem

\bibitem[\protect\citeauthoryear{Tautz}{2012}]{tau12:nov}
\begin{bchapter}
\bauthor{\bsnm{Tautz}, \binits{R.C.}}:
In: \beditor{\bsnm{Marcuso}, \binits{R.J.}} (ed.)
\bbtitle{{Turbulence: Theory, Types and Simulation}},
p. \bfpage{365}.
\bpublisher{Nova Publishers},
\blocation{New York}
(\byear{2012})
\end{bchapter}
\endbibitem

\bibitem[\protect\citeauthoryear{Tautz and Lerche}{2012a}]{tau12:rad}
\begin{barticle}
\bauthor{\bsnm{Tautz}, \binits{R.C.}},
\bauthor{\bsnm{Lerche}, \binits{I.}}:
\bjtitle{Phys. Rep.}
\bvolume{520},
\bfpage{1}
(\byear{2012}a)
\end{barticle}
\endbibitem

\bibitem[\protect\citeauthoryear{Tautz and Lerche}{2012b}]{tau12:3x3}
\begin{barticle}
\bauthor{\bsnm{Tautz}, \binits{R.C.}},
\bauthor{\bsnm{Lerche}, \binits{I.}}:
\bjtitle{J. Math. Phys.}
\bvolume{53},
\bfpage{083302}
(\byear{2012}b)
\end{barticle}
\endbibitem

\bibitem[\protect\citeauthoryear{Tautz and Sakai}{2007}]{tau07:amp}
\begin{barticle}
\bauthor{\bsnm{Tautz}, \binits{R.C.}},
\bauthor{\bsnm{Sakai}, \binits{J.-I.}}:
\bjtitle{Phys. Plasmas}
\bvolume{14},
\bfpage{012104}
(\byear{2007})
\end{barticle}
\endbibitem

\bibitem[\protect\citeauthoryear{Tautz and Sakai}{2008}]{tau08:ba1}
\begin{barticle}
\bauthor{\bsnm{Tautz}, \binits{R.C.}},
\bauthor{\bsnm{Sakai}, \binits{J.-I.}}:
\bjtitle{J. Plasma Phys.}
\bvolume{74},
\bfpage{79}
(\byear{2008})
\end{barticle}
\endbibitem

\bibitem[\protect\citeauthoryear{Tautz and Schlickeiser}{2005a}]{tau05:co1}
\begin{barticle}
\bauthor{\bsnm{Tautz}, \binits{R.C.}},
\bauthor{\bsnm{Schlickeiser}, \binits{R.}}:
\bjtitle{Phys. Plasmas}
\bvolume{12},
\bfpage{122901}
(\byear{2005}a)
\end{barticle}
\endbibitem

\bibitem[\protect\citeauthoryear{Tautz and Schlickeiser}{2005b}]{tau05:cov}
\begin{barticle}
\bauthor{\bsnm{Tautz}, \binits{R.C.}},
\bauthor{\bsnm{Schlickeiser}, \binits{R.}}:
\bjtitle{Phys. Plasmas}
\bvolume{12},
\bfpage{072101}
(\byear{2005}b)
\end{barticle}
\endbibitem

\bibitem[\protect\citeauthoryear{Tautz and Schlickeiser}{2006}]{tau06:co2}
\begin{barticle}
\bauthor{\bsnm{Tautz}, \binits{R.C.}},
\bauthor{\bsnm{Schlickeiser}, \binits{R.}}:
\bjtitle{Phys. Plasmas}
\bvolume{13},
\bfpage{062901}
(\byear{2006})
\end{barticle}
\endbibitem

\bibitem[\protect\citeauthoryear{Tautz and Schlickeiser}{2007}]{tau07:spo}
\begin{barticle}
\bauthor{\bsnm{Tautz}, \binits{R.C.}},
\bauthor{\bsnm{Schlickeiser}, \binits{R.}}:
\bjtitle{Phys. Plasmas}
\bvolume{14},
\bfpage{102102}
(\byear{2007})
\end{barticle}
\endbibitem

\bibitem[\protect\citeauthoryear{Tautz and Shalchi}{2008}]{tau08:wei}
\begin{barticle}
\bauthor{\bsnm{Tautz}, \binits{R.C.}},
\bauthor{\bsnm{Shalchi}, \binits{A.}}:
\bjtitle{Phys. Plasmas}
\bvolume{15},
\bfpage{052304}
(\byear{2008})
\end{barticle}
\endbibitem

\bibitem[\protect\citeauthoryear{Tautz et~al.}{2006}]{tau06:har}
\begin{barticle}
\bauthor{\bsnm{Tautz}, \binits{R.C.}},
\bauthor{\bsnm{Lerche}, \binits{I.}},
\bauthor{\bsnm{Schlickeiser}, \binits{R.}}:
\bjtitle{Phys. Plasmas}
\bvolume{13},
\bfpage{052112}
(\byear{2006})
\end{barticle}
\endbibitem

\bibitem[\protect\citeauthoryear{Tautz et~al.}{2007}]{tau07:har}
\begin{barticle}
\bauthor{\bsnm{Tautz}, \binits{R.C.}},
\bauthor{\bsnm{Lerche}, \binits{I.}},
\bauthor{\bsnm{Schlickeiser}, \binits{R.}}:
\bjtitle{J. Math. Phys.}
\bvolume{48},
\bfpage{013302}
(\byear{2007})
\end{barticle}
\endbibitem

\bibitem[\protect\citeauthoryear{Tautz et~al.}{2006}]{tau06:is1}
\begin{barticle}
\bauthor{\bsnm{Tautz}, \binits{R.C.}},
\bauthor{\bsnm{Lerche}, \binits{I.}},
\bauthor{\bsnm{Schlickeiser}, \binits{R.}},
\bauthor{\bsnm{Schaefer-Rolffs}, \binits{U.}}:
\bjtitle{J. Phys.~A: Math. Gen.}
\bvolume{39},
\bfpage{13831}
(\byear{2006})
\end{barticle}
\endbibitem

\bibitem[\protect\citeauthoryear{Weibel}{1959}]{wei59:wei}
\begin{barticle}
\bauthor{\bsnm{Weibel}, \binits{E.S.}}:
\bjtitle{Phys. Rev. Lett.}
\bvolume{2},
\bfpage{83}
(\byear{1959})
\end{barticle}
\endbibitem

\bibitem[\protect\citeauthoryear{Yoon}{2007}]{yoo07:spo}
\begin{barticle}
\bauthor{\bsnm{Yoon}, \binits{P.H.}}:
\bjtitle{Phys. Plasmas}
\bvolume{14},
\bfpage{064504}
(\byear{2007})
\end{barticle}
\endbibitem

\bibitem[\protect\citeauthoryear{Yoon and Schlickeiser}{2012}]{yoo12:sp1}
\begin{barticle}
\bauthor{\bsnm{Yoon}, \binits{P.H.}},
\bauthor{\bsnm{Schlickeiser}, \binits{R.}}:
\bjtitle{Phys. Plasmas}
\bvolume{19},
\bfpage{022105}
(\byear{2012})
\end{barticle}
\endbibitem

\bibitem[\protect\citeauthoryear{Zaheer and Murtaza}{2007}]{zah07:sem}
\begin{barticle}
\bauthor{\bsnm{Zaheer}, \binits{G.}},
\bauthor{\bsnm{Murtaza}, \binits{G.}}:
\bjtitle{Phys. Plasmas}
\bvolume{14},
\bfpage{072106}
(\byear{2007})
\end{barticle}
\endbibitem

\bibitem[\protect\citeauthoryear{Zweibel and Shull}{1982}]{zwe82:mol}
\begin{barticle}
\bauthor{\bsnm{Zweibel}, \binits{E.G.}},
\bauthor{\bsnm{Shull}, \binits{J.M.}}:
\bjtitle{Astrophys. J.}
\bvolume{259},
\bfpage{859}
(\byear{1982})
\end{barticle}
\endbibitem

\end{thebibliography}

\end{document}